\documentclass[twocolumn,prc,preprintnumbers,superscriptaddress,floatfix,showpacs]{revtex4}
\usepackage{dcolumn}
\usepackage{bm}
\usepackage{longtable}
\usepackage{graphicx,epsfig,latexsym,amssymb}
\usepackage{multirow,amsmath,array,booktabs,color}
\usepackage[section]{placeins}

\begin{document}

\title{Combination of complex momentum representation and Green's function methods in relativistic mean-field theory}

\author{Min Shi}
\affiliation{School of mathematics and physics, Anhui Jianzhu University, Hefei 230601, People's Republic of China}

\author{Zhong-Ming Niu}
\email{zmniu@ahu.edu.cn}
\affiliation{School of physics and materials science, Anhui University, Hefei 230601, People's Republic of China}

\author{Haozhao Liang}
\email{haozhao.liang@riken.jp}
\affiliation{RIKEN Nishina Center, Wako 351-0198, Japan}
\affiliation{Department of Physics, Graduate School of Science,
The University of Tokyo, Tokyo 113-0033, Japan}

\date{\today }

\begin{abstract}
We have combined the complex momentum representation method with the Green's function method in the relativistic mean-field framework to establish the RMF-CMR-GF approach. This new approach is applied to study the halo structure of $^{74}$Ca. All the continuum level density of concerned resonant states are calculated accurately without introducing any unphysical parameters, and they are independent of the choice of integral contour. The important single-particle wave functions and densities for the halo phenomenon in $^{74}$Ca are discussed in detail.
\end{abstract}


\maketitle

\section{Introduction}

The exotic nuclei near the drip-lines are generally loosely bound or unbound, and nowadays the study of exotic nuclei has achieved more and more attention with the development of radioactive nuclear beam experiments \cite{Tanihata1996}. The Fermi surfaces of these nuclear systems are usually close to the single-particle resonances. As a result, the resonances play crucial roles in forming various exotic nuclear phenomena \cite{Dobaczewski1996, Poschl1997, Meng1996, Zhang2012, Zhou2010, Jensen2004, Meng2006, Meng2015}. Therefore, the proper treatment of resonances is crucial for understanding the phenomena in exotic nuclei, such as halo, giant halo, deformed halo, and so on.

Many methods have been proposed to study the resonances. On the one hand, based on the conventional scattering theory, the $\emph{R}$-matrix \cite{Wigner1947,Hale1987}, $\emph{K}$-matrix \cite{Humblet1991}, and $\emph{S}$-matrix \cite{Taylor1972} theories have been developed. On the other hand, many bound-state-like methods have been developed as well, in which the resonant states can be treated similarly to the bound states. For example, the resonance parameters in the real stabilization method (RSM) can be calculated from the change in the stable eigenvalue with the increase of basis size \cite{Hazi1970}. The analytic continuation in the coupling constant (ACCC) method regards a resonance as an analytic continuation of a bound state \cite{Kukulin1989}. The complex energy eigenvalues can be discretized for the bound, resonant, and non-resonant continuum states by diagonalizing the complex-scaled Hamiltonian matrix in the complex scaling method (CSM) \cite{Aguilar1971, Balslev1971, Simon1972, Myo2014}. In addition, the resonances can be also probed in the complex momentum space \cite{Sukumar1979, Kwon1978, Berggren1968}.

Due to the success of the relativistic mean-field (RMF) theory in describing the exotic properties in nuclei \cite{Walecka1974, Serot1986, Ring1996, Vretenar2005, Liang2015, Liang2008, Niu2013a, Niu2017,Paar2009} and its application in nuclear astrophysics \cite{Sun2008, Niu2009, Xu2013, Niu2013, Niu2011}, the methods mentioned above have been combined with the RMF theory, including the RMF-RSM approach \cite{Zhang2008}, the RMF-ACCC approach \cite{Yang2001, Zhang2004}, and the RMF-CSM approach \cite{Guo2005}.

Meanwhile, the study of continuum level density (CLD) also catches a wide attention, which is another way to probe the resonances. The CLD has been calculated in the nonrelativistic framework \cite{Suzuki2005}. Using the complex scaled Green's function method, the study of CLD has been extended to the relativistic framework in the spherical and deformed nuclei \cite{Shi2014, Shi2017}. Although such methods can describe the resonant states in an intuitive way, their results are not completely independent on the rotation angle in the actual calculations with a finite basis.

Recently, Guo $\emph{et al.}$ applied the complex momentum representation (CMR) method to the RMF theory and developed the RMF-CMR approach. This approach has already been applied to not only the spherical case~\cite{Li2016} but also the deformed case \cite{Fang2017}. By using the RMF-CMR approach, both the bound states and resonant states can be treated at the same time, and the self-consistent results can be obtained. It is worthwhile to mention that not only this method can be reliably applied to narrow resonances, but also it is very effective for the broad resonances, which cannot be well described by many other bound-state-like methods.

Therefore, it is meaningful to combine the complex momentum representation method and the Green's function method in the relativistic mean-field framework to establish the so-called RMF-CMR-GF method, which can obtain the CLD accurately and efficiently yet without introducing any unphysical parameters.

Since there are rich experimental data as well as unsolved mysteries on various properties of Calcium (Ca) isotopes, the studies on the Ca isotopes have attracted a wide attention in recent years. The halos in the Ca isotopes near the drip-line have been studied by the relativistic continuum Hartree-Bogoliubov theory \cite{Meng2002}. In addition, based on the relativistic Hartree-Fock-Bogoliubov theory \cite{Long2010}, the halo structures in the Ca isotopes were further studied in detail, and distinct evidence of halo occurrence was found, in which the single-particle states around the particle continuum threshold play important roles in forming the halos. In these halo nuclei, the single-particle resonant states are generally near the continuum threshold, which has been studied systematically in the Ca isotopes by the RMF-RSM approach \cite{Zhang2010}.

In this work, we will employ the RMF-CMR-GF method to investigate the resonances via the continuum level density. The RMF-CMR-GF method will be developed and the basic formula will be given in Section~\ref{Sec:Formalism}. By taking the halo nucleus $^{74}$Ca as an example, the results and discussion will be given in Section~\ref{Sec:Results}. A summary will be exhibited in Section~\ref{Sec:Summary}.

\section{Formalism}\label{Sec:Formalism}

In the momentum representation, the Dirac equation can be expressed as \cite{Li2016}
\begin{equation}\label{Diraceq2}
\int d\vec{k}^{\prime }\langle \vec{k}\vert H\vert \vec{k}^{\prime }\rangle \psi ( \vec{k}^{\prime }) =\varepsilon
\psi ( \vec{k}),
\end{equation}%
where $\vert \vec{k}\rangle $ is the wave function of a free particle with momentum $\vec{p}$ or wave vector $\vec{k} = \vec{p}/\hbar $, $H = \vec{\alpha}\cdot\vec{p}+\beta \left( M+S\right) +V$ is the Dirac Hamiltonian, and $\psi( \vec{k}) $ is the momentum wave function.

For a spherical system, $\psi(\vec{k})$ can be split into the radial and angular parts as
\begin{equation}\label{wave function}
\psi ( \vec{k}) =\left(
\begin{array}{c}
f(k)\phi _{ljm_{j}}\left( \Omega _{k}\right) \\
g(k)\phi _{\tilde{l}jm_{j}}\left( \Omega _{k}\right)
\end{array}
\right) ,
\end{equation}
where $\phi _{ljm_{j}}\left(\Omega _{k}\right)$ is a two-component spinor. The quantum number of the orbital angular momentum corresponding to the upper (lower) component of Dirac spinor is denoted as $l$ ($\tilde{l}$). The relationship between these two quantum numbers and the total angular momentum quantum number $j$ reads $\tilde{l}=2j-l$.

Putting Eq.~(\ref{wave function}) into Eq.~(\ref{Diraceq2}) and turning an integral into a sum over a finite set of points $k_{j}$ and $dk$ with a set of weights $w_{j}$ \cite{Li2016}, we get a symmetric matrix equation,
\begin{equation}
\sum\limits_{j=1}^{N}\left(
\begin{array}{cc}
A_{ij}^{+} & B_{ij} \\
B_{ij} & A_{ij}^{-}%
\end{array}%
\right) \left(
\begin{array}{c}
\mathfrak{f}(k_{j}) \\
\mathfrak{g}(k_{j})%
\end{array}%
\right) =\varepsilon \left(
\begin{array}{c}
\mathfrak{f}(k_{i}) \\
\mathfrak{g}(k_{i})%
\end{array}%
\right) ,
\end{equation}
with
\begin{subequations}
\begin{align}
V^{+}\left( k,k^{\prime }\right) & =\frac{2}{\pi }\int r^{2}dr\left[ V\left(
r\right) +S\left( r\right) \right] j_{l}\left( k^{\prime }r\right)
j_{l}\left( kr\right) , \\
V^{-}\left( k,k^{\prime }\right) & =\frac{2}{\pi }\int r^{2}dr\left[ V\left(
r\right) -S\left( r\right) \right] j_{\tilde{l}}\left( k^{\prime }r\right)
j_{\tilde{l}}\left( kr\right).
\end{align}
\end{subequations}
Here, $A_{ij}^{\pm }={\pm }M\delta _{ij}+\sqrt{w_{i}w_{j}} k_{i}k_{j}V^{\pm }\left( k_{i},k_{j}\right) $ and $B_{ij}=-k_{i}\delta _{ij}$. The spherical Bessel functions of orders $l$ are $\tilde{l}$ are denoted as $j_{l}(kr)$ and $j_{\tilde{l}}(kr)$, respectively. In addition, $\mathfrak{f}(k)$ and $\mathfrak{g}(k)$ are the symmetric forms of the radial parts of Dirac spinor. Now, the problem of solving the Dirac equation becomes a problem of solving the symmetric matrix. In actual calculations, after we choose a proper contour for the momentum integration, the bound states populate on the imaginary axis in the momentum plane, while the resonances locate at the fourth quadrant. The upper and lower components of wave functions in the coordinate space can be obtained with
\begin{subequations}
\begin{align}
f(r) &=i^{l}\sqrt{\frac{2}{\pi }}\sum\limits_{j=1}^{N}\sqrt{w_{j}}%
k_{j}j_{l}(k_{j}r)\mathfrak{f}(k_{j}), \\
g(r) &=i^{l}\sqrt{\frac{2}{\pi }}\sum\limits_{j=1}^{N}\sqrt{w_{j}}%
k_{j}j_{l}(k_{j}r)\mathfrak{g}(k_{j}).
\end{align}
\end{subequations}

Similar to Ref.~\cite{Myo2014}, the level density in the complex momentum Green's function can be expressed as
\begin{align}
\rho (\varepsilon ) =-\frac{1}{\pi }\mathrm{Im}\int {d\mathbf{r}}\Bigg[ & %
\sum_{{b}}^{N{_{b}}}\frac{\psi _{b}({\mathbf{r}})\tilde{\psi}{_{b}^{\ast }(%
\mathbf{r})}}{\varepsilon -\varepsilon _{b}}  \notag \\
&+\sum_{{r}}^{N_{r}}\frac{\psi _{r}({\mathbf{r}})\tilde{\psi}{_{r}^{\ast }(%
\mathbf{r})}}{\varepsilon -\varepsilon _{r}}  \notag \\
&+\int d\varepsilon _{c}\frac{\psi _{c}({\mathbf{r}})\tilde{\psi}_{c}^{\ast
}{(\mathbf{r})}}{\varepsilon -\varepsilon _{c}}\Bigg],  \label{density}
\end{align}
where $\psi _{b}({\mathbf{r}})$, $\psi _{r}({\mathbf{r}})$, and $\psi _{c}({\mathbf{r}})$ are the wave functions for the bound states, resonant states, and the non-resonance continuum in coordinate space, and $\tilde{\psi}^{\ast }({\mathbf{r}})$ is the Hermite conjugate of the corresponding wave function. Meanwhile, $\varepsilon _{b},$ $\varepsilon _{r},$ and $\varepsilon _{c}$ are the energy eigenvalues for the bound states, resonant states, and the non-resonance continuum, respectively. Similar to Refs.~\cite{Suzuki2005,Shi2014,Shi2017,Myo2014}, we can further get the CLD, denoted as $\Delta \rho (\varepsilon )$, by the difference between the density of states $\rho (\varepsilon )$ and the density of continuum states $\rho _{0}(\varepsilon )$. Here, $\rho_{0}(\varepsilon )$ is obtained from the asymptotic Hamiltonian $H_{0}$ in the form of $H$ with $r\rightarrow \infty $.

\section{Results and Discussion}\label{Sec:Results}

\begin{figure}
\includegraphics[width=8cm]{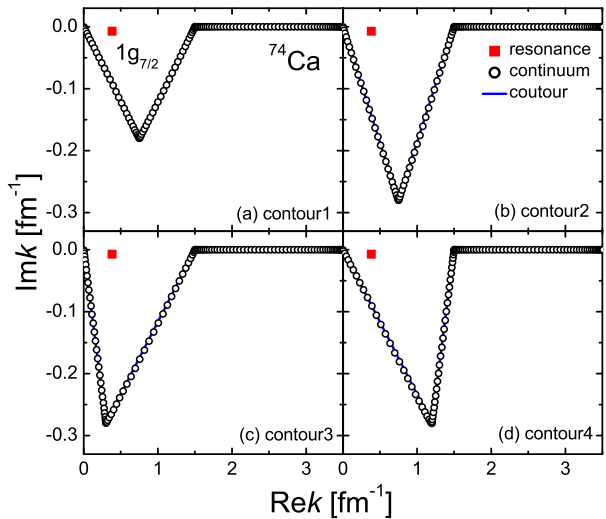}
\caption{(Color online) Single-particle resonances for the $g_{7/2}$ orbitals in $^{74}$Ca calculated by the RMF-CMR-GF approach with the interaction NL3 in four different contours for the momentum integration. The red solid squares, black circles, blue solid line represent the resonant state $1g_{7/2}$, the continuum, and the contour of integration in the complex momentum plane, respectively.}
\label{Fig1}
\end{figure}

In Fig.~\ref{Fig1}, we show the single-particle resonant state $1g_{7/2}$ in $^{74}$Ca calculated by the RMF-CMR-GF approach with the interaction NL3 in four different contours. As shown in the figure, the contour1 is characterized by four points in the complex momentum plane, $k_1 = 0$~fm$^{-1}$, $k_2 = 0.75 -i0.18$~fm$^{-1}$, $k_3 = 1.5$~fm$^{-1}$, and $k_{\text{max}} = 3.5$~fm$^{-1}$. Three points out of four in the other contours are the same as those in the contour1, but $k_2 = 0.75 -i0.28$~fm$^{-1}$, $0.30 -i0.28$~fm$^{-1}$, and $1.20 -i0.28$~fm$^{-1}$ for the contour2, contour3, and contour4, respectively. In each panel, it can be seen that the resonant state has been exposed clearly in the complex momentum plane. No matter the contour becomes deeper from contour1 to contour2, or the contour moves from left to right from contour3 to contour4, the physical resonant state $1g_{7/2}$ always keeps a fixed position, while the corresponding continuous spectra follow the varied contours.

\begin{figure}
\includegraphics[width=8cm]{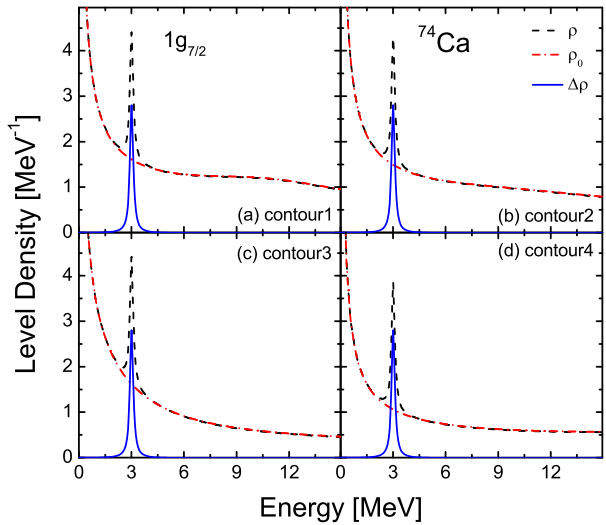}
\caption{(Color online) Density of states $\rho(\varepsilon)$, density of continuum states $\rho_{0}(\varepsilon)$, and CLD $\Delta\rho(\varepsilon)$ for the $g_{7/2}$ orbitals in $^{74}$Ca by the RMF-CMR-GF calculations with four different contours. These level densities are denoted with the black dashed, red dash-dotted, and blue solid lines, respectively.}
\label{Fig2}
\end{figure}

In Fig.~\ref{Fig2}, the density of states $\rho(\varepsilon)$, the density of continuum states $\rho_{0}(\varepsilon)$, and the CLD $\Delta\rho(\varepsilon)$ for the $g_{7/2}$ orbitals in $^{74}$Ca are shown, with the four contours that are the same as those in Fig.~\ref{Fig1}. The density of states $\rho(\varepsilon)$ corresponds to the level density of the total Hamiltonian $\emph{H}$. It includes the resonant states and non-resonance continuum, and shows a very sharp peak. In contrast, the density of continuum states $\rho_{0}(\varepsilon)$ is obtained from the asymptotic Hamiltonian $\emph{H}_{0}$, and it shows no peak because there is no centrifugal barrier in the potential of the asymptotic Hamiltonian. For that, $\rho_{0}(\varepsilon)$ is often called the density of background energy. Finally, the CLD $\Delta\rho(\varepsilon)$ can be obtained by subtracting $\rho_{0}(\varepsilon)$ from $\rho(\varepsilon)$. Its sharp peak corresponding to the resonance can be shown clearly.

As the momentum integration can be carried out with different contours in the RMF-CMR-GF calculations, it is necessary to check whether the CLD varies with the contours. Comparing with different panels in Fig.~\ref{Fig2}, it is seen that although $\rho(\varepsilon)$ and $\rho_{0}(\varepsilon)$ change with different contours, the CLD $\Delta\rho(\varepsilon)$ always keeps its position, height and width. This demonstrates that the resonance parameters can be determined independently on the contour in the complex momentum plane, and the present RMF-CMR-GF approach is efficient in describing the resonances of a Dirac particle.

\begin{figure}
\includegraphics[width=8.5cm]{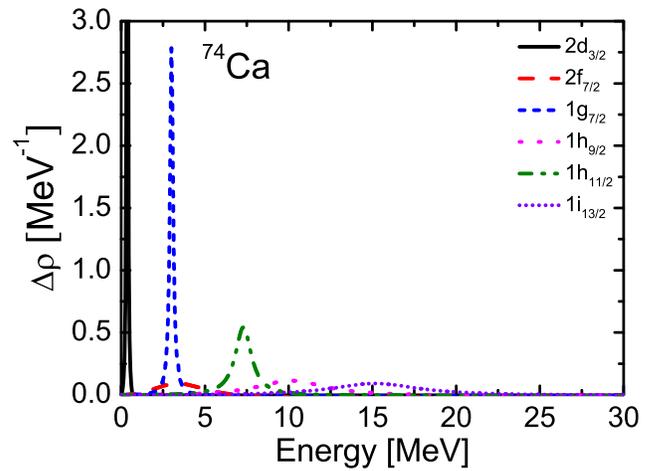}
\caption{(Color online) The CLD $\Delta\rho(\varepsilon)$ for all the resonant states in $^{74}$Ca obtained by the RMF-CMR-GF calculations, where the contour2 is adopted.}
\label{Fig3}
\end{figure}

Since the continuum level density is independent on the integral contour, all the concerned resonant states in $^{74}$Ca can be obtained as long as the range of contour is large enough. The corresponding results are shown in Fig.~\ref{Fig3}. It is noted that the density of the non-resonance continuum has been subtracted. As a result, the resonance peaks can be exhibited explicitly and the resonances with different angular momentum have different positions and widths. The position of resonance peak corresponds to the energy of the resonant state, and the width of the resonance peak at half height corresponds the width of the resonant state. The lifetime of the resonance is inversely proportional to its width, which indicates, for example, the resonant state $1i_{13/2}$ with a broad peak has a short lifetime, in comparison, the narrow resonances $2d_{3/2}$ and $1g_{7/2}$ have longer lifetimes.

\begin{figure}
\includegraphics[width=8.5cm]{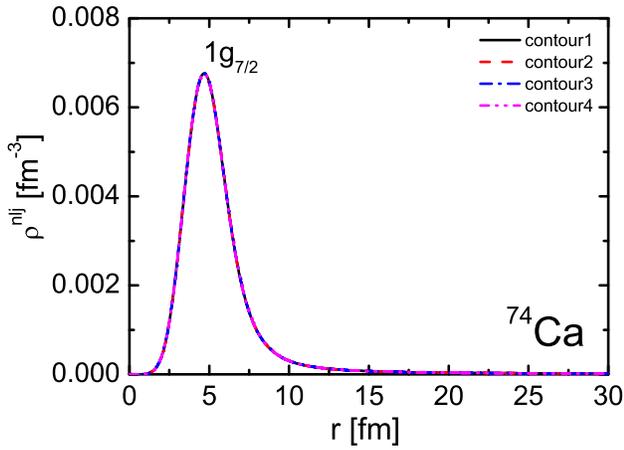}
\caption{(Color online) Density of resonant state $1g_{7/2}$ in coordinate space with four different contours for the momentum integration in the RMF-CMR-GF calculations, where the black solid, red dashed, blue dash-dotted, and magenta dash-dot-dotted lines represent the contour1, contour2, contour3, and contour4, respectively.}
\label{Fig4}
\end{figure}

Apart from the CLD, we can also obtain the density of different orbitals in $^{74}$Ca in the coordinate space. The densities of the resonant state $1g_{7/2}$ with four different contours are drawn in Fig.~\ref{Fig4}. We can see that the density of the resonant state is also independent on the choice of the integral contour.

\begin{figure}
\includegraphics[width=8.5cm]{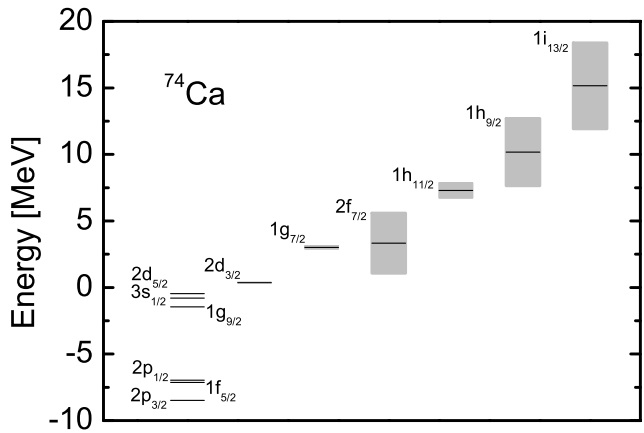}
\caption{Single-particle spectra in $^{_{74}}$Ca in the RMF-CMR-GF calculations with the interaction NL3. Heights of the gray rectangles represent the widths of corresponding resonances.}
\label{Fig5}
\end{figure}

After we examine the applicability and efficiency of the RMF-CMR-GF method in describing the resonant states, all the concerned bound states and resonant states in $^{74}$Ca are obtained. Figure~\ref{Fig5} shows the bound and resonant states with the energies between $-10$ and $20$~MeV, where the contour2 is adopted. In $^{74}$Ca the Fermi surface locates at the last bound state $2d_{5/2}$. The resonant state $2d_{3/2}$ located near the particle continuum threshold will be populated with pairing correlation, which forms the halo phenomenon~\cite{Meng2002}.

\begin{figure}
\includegraphics[width=8.5cm]{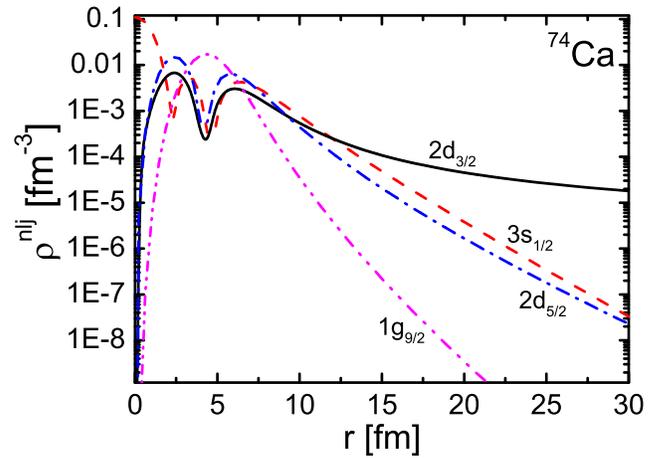}
\caption{(Color online) Density in logarithm scale for the orbitals $2d_{3/2}$, $3s_{1/2}$, $2d_{5/2}$, and $1g_{9/2}$. They are labeled as the black solid, red dashed, blue dash-dotted, and magenta dash-dot-dotted lines, respectively.}
\label{Fig6}
\end{figure}

In order to further analyze the mechanism for forming the halo structure of $^{74}$Ca, we show the densities of different orbitals in Fig.~\ref{Fig6}. With the increasing radius $r$, the densities of the bound states $3s_{1/2}$, $2d_{5/2}$, and $1g_{9/2}$ exponentially decay to zero, while the density of the resonant state $2d_{3/2}$ has a long tail. This resonant state is crucial for forming the halo structure in $^{74}$Ca.

\begin{figure}
\includegraphics[width=8.5cm]{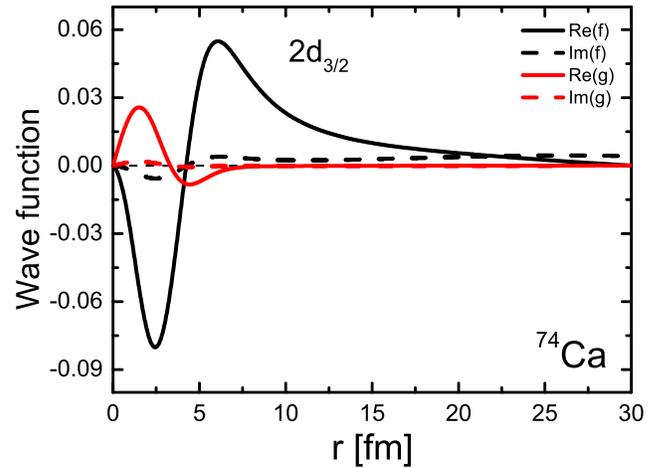}
\caption{(Color online) Wave functions of the resonant state $2d_{3/2}$. The black (red) solid and dashed lines represent the real and imaginary parts of the upper (lower) component $f(r)$ ($g(r)$).}
\label{Fig7}
\end{figure}

To further explain the reason why the density of the resonant state $2d_{3/2}$ decreases gently in the region with a large radius, we show its wave functions in Fig.~\ref{Fig7}, with both the real and imaginary parts of the upper and lower components of Dirac spinor. It is seen that the contribution to the density mainly comes from the upper component of the wave function. The real part of the upper component $f(r)$ has a tail at a large radius, while the real part of the lower component $g(r)$ decreases quickly to $0$. Meanwhile, it is worth pointing out that the imaginary part of the upper component $f(r)$ increases with the increasing radius, which is another key contribution to the density.

\section{Summary}\label{Sec:Summary}

We have established the RMF-CMR-GF approach, and the theoretical formalism has been presented in this paper. We have obtained the bound and resonant states on the same footing in the momentum space. All the concerned resonant states of $^{74}$Ca are obtained, including not only the narrow resonances but also the broad resonances. By subtracting the density of continuum states $\rho_{0}(\varepsilon)$ from the density of states $\rho(\varepsilon)$, the CLD $\Delta\rho(\varepsilon)$ is obtained, which is independent on the choice of the integral contour. The resonant peaks can be exposed clearly, and the resonance energies and widths can be determined in an accurate way. The densities of different orbitals are shown to distinguish the halo orbitals, and the wave functions of the resonant state $2d_{3/2}$ are exhibited to further explain the halo structure of $^{74}$Ca. The present results show the applicability and efficiency of the RMF-CMR-GF method for the study of resonant states in nuclei, in particular, the halo structure in exotic nuclei.

\section*{Acknowledgments}

This work was partly supported by the National Natural Science Foundation of China under Grant No.~11205004, the Natural Science Foundation of Anhui Province under Grant No.~1708085QA10, the Key Research Foundation of Education Ministry of Anhui Province under Grant No.~KJ2016A026, the RIKEN iTHES project and iTHEMS program, and the Doctor Foundation of Anhui Jianzhu University 2017 under Grant No.~2017QD18.

\end{document}